\documentclass[3p,number]{elsarticle}

\usepackage{amsmath,amssymb}
\usepackage{graphicx,amsfonts,graphicx,multirow,color}
\usepackage{hyperref}

\expandafter\ifx\csname package@font\endcsname\relax\else
 \expandafter\expandafter
 \expandafter\usepackage
 \expandafter\expandafter
 \expandafter{\csname package@font\endcsname}%
\fi

\newcommand{\DD}[2]{\frac{\partial #1}{\partial #2}}

\newcommand{\rd}{\mathrm{d}}

\begin{document}
\title{On the Pseudolocalized Solutions in Multi-dimension of Boussinesq Equation}
\author{C. I. Christov}
\ead{christov@louisiana.edu}
\address{Department of Mathematics, University of  Louisiana at  Lafayette,
Lafayette, LA 70504, USA}
\date{\today}

\begin{abstract}
A new class of solutions of  three-dimensional equations from the Boussinesq paradigm are considered. The corresponding profiles are not localized functions in the sense of the integrability of the square over an infinite domain. For the new type of solutions, the gradient and the Hessian/Laplacian  are square integrable. In the linear limiting case, analytical expressions for the profiles of the pseudolocalized solutions are found. The nonlinear case is treated numerically with a special approximation of the differential operators with spherical symmetry that allows for automatic acknowledgement of the behavioral conditions at the origin of the coordinate system. The asymptotic boundary conditions stem from the $1/r$ behavior at infinity of the pseudolocalized profile. A special approximation is devised that allows us to obtain the proper behavior for much smaller computational box. The pseudolocalized solutions are obtained for both quadratic and cubic nonlinearity.

\end{abstract}

%\pacs{44.05.+e, 44.10.+i,05.60.-k}

\begin{keyword}{Pseudolocalized Solutions, Quasi-Particles, Peakons}
\end{keyword}

\maketitle

%(Least Action Principle for Dissipative Systems)

\section{Introduction: Localized Waves and Quasi-Paritcles}

John Scott-Russell \cite{Russ38} observed the permanent (or ``great'') wave that can travel large distances without changing form.  In 1871-1872 Boussinesq came up with a fundamental idea: the balance between the nonlinearity and dispersion is what makes the shape of the wave permanent \cite{Boussi71a,Boussi72}.  He found the first analytical solution of $sech$ type. Later on, Korteweg and de Vries \cite{Kort95}  showed that in a frame moving with the characteristic speed, the second order in time Boussinesq equation can be reduced to a first-order in time equation (known now as KdV equation)  for which the balance between nonlinearity and dispersion holds as it does for its parent: the Boussinesq equation. Boussinesq-type equations arise not only in shallow water flows and also in the theory of shells and plates (the famous von K\'arm\'an equations) \cite{Karman,Ciarlet_vonKaman}. Even the Schr\"odinger equation of wave mechanics  can be shown to be a Boussinesq equation for the real or imaginary part of the wave function \cite{Lumini_WCNA}.
Boussinesq equation is a generalized wave equation containing dispersion in the form of a biharmonic operator of the sought function. The dispersive effects of the biharmonic operator can be countered by the presence of a nonlinearity, and as a result, a permanent wave of localized type may exist and propagate without change. We call this balance between the nonlinearity and dispersion the ``Boussinesq paradigm".

One of the main properties of the ``permanent'' localized waves is the collision property. Skyrme \cite{Skyrme61} argued the idea that different particles are the localized waves of different field equations and actually found a particle-like behavior of the localized solution of the $sine$-Gordon equation \cite{Perring_skyrme}. Unfortunately, the latter solution were kinks (`hydraulic jumps') which did not appeal directly the physical intuition of particles as being localized `lumps' of energy. Actually the works of Skyrme defined in essence the notion of Soliton which was introduced a couple of year later by  Zabusky and Kruskal \cite{Zabu65} who discovered numerically that the solitary waves of KdV equation retain their shapes after multiple collisions. They introduced the coinage \emph{soliton} to emphasize this particle-like behavior. Currently the term \emph{soliton} in strict sense is reserved only for the particle-like localized solution of fully integrable systems.  For integrable and nonintegrable systems alike, the evolution of a wave system which is a superposition of two localized waves results in a virtual recovery of the shapes of the two initial localized waves but in a shifted positions: outgoing from the site of interaction. The collision property gave the rise of the term ``quasi-particle" (or QP). When the physical system is described by equation(s) that conserve the energy and momentum and some integral interpreted as the pseudomass of a localized shape, the quasi-particles behave very much as actual particles in quantum physics.

Yet there are essential differences between the quasi-particles currently known and the real particles. The most conspicuous is that the research is mostly concerned with 1D cases, which have virtually no relevance to the physical reality. Even within the class of 1D QPs there are more questions to be answered before they are shown to qualify as particles. One of the unsatisfactory features is that the quasi-particles actually pass through each other (in a sense, they `percolate', in a sense, through each other) rather then scatter. This does not prevent the investigators to speak about scattering of solitons. To the limit of author's knowledge, the only work that attempts to answer the question of whether the $sine$-Gordon kinks percolate or scatter is \cite{FerguWillis,SG_AIMS}, where the Variational Approximation is applied. The standard approach without additional collective variables (see \cite{IvanDad_sG}, and literature cited therein) cannot answer this question.

It has been just recently discovered (see \cite{repel_AIMS} and the further corroboration in  \cite{Josh_WM})  that under certain conditions the solitons of the System of Coupled Nonlinear Schr\"odinger Equations (SCNLSE) (called alternatively Vector NLSE) can actually scatter without crossing each other (percolating through each other).

\subsection{Classical Solitons as Quasi-Particles: The Con's}

Mathematicians place emphasis on the full integrability of the system that exhibits soliton solutions. However, physically speaking, the three laws known to govern motion are the conservation of mass, energy, and momentum. The full integrability is not directly relevant for the physical implication of localized solutions as quasi-particles. 

There are essential differences between the real particles and 1D quasi-particles known currently for which hundreds of strict mathematical results are proven. \emph{First}, the known quasi-particles actually pass through each other (percolate in a sense), i.e. they do not repel (scatter) like the real particles. \emph{Second}, the repelling or attracting of the QPs depend very much on the potential of interaction defined by the asymptotic behavior of their tails. In 1D, for virtually all main soliton supporting equations the potential is attractive, and decays exponentially with the distance between the QPs. In real-world physics, at large distances the potential (producing the gravitation force) decays algebraically as the inverse of the distance, which is a radical difference from an exponential decay. \emph{Third}, the results about the interaction in more than one dimension are rare and mostly qualitative, not quantitative. The repelling or attracting of the QPs depend very much on the potential of interaction defined by the asymptotic behavior of their tails. 

The above list of discrepancies between the currently known QPs and the actual physical particles can be extended, but the main points are those mentioned above. 
\subsection{The First Forays in the Realm of 2D Boussinesq Solitons}

The first thing that comes to mind is that this exponentially decaying asymptotic behavior is due to the one-dimensional nature of known QPs. Unfortunately, very little is known for solitary waves that are truly localized in the two spatial variable. For the time being, the main success has been in finding solitary waves of the Kadomtsev--Petviashvily (KP) equation (see the original article \cite{KadomPetvi}, also \cite{ChriMauPorub,PorubMauMateev,PorubPastronMau} and the literature cited therein).

The strictly localized steadily propagating solutions of the truly 2D equations of Boussinesq type were not investigated until recently. The question is very important and we have developed three different numerical techniques to interrogate the problem. A very fast Galerkin spectral method is used in  \cite{Athens_2DBoussi}; a semi-analytical method for relatively small propagation speed is presented in \cite{ChriChou_asymp}; and a finite-difference solution with special implementation of the asymptotic boundary condition is given in \cite{AsympNum_2D}. The three methods agree quantitatively very well in the common domains of the values of the parameters, so that the findings can be considered validated with high accuracy.

The findings about the 2D localized solution of Boussinesq equation can be summarized as follows. The profile of the standing soliton has a super exponential decay at infinity, namely $e^{-r}/\sqrt{r}$. This property is not robust, and even for very small propagation speeds, the decay at infinity changes to algebraic, proportional to $1/r^2$. However, the most disappointing property is the fact that the steadily propagating 2D shapes are not stable and after some time they either dissipate or blow up. This important result was first found by means of a special new scheme \cite{ChertChriKurgan} and confirmed quantitatively in \cite{ChriKolkVasi} with a different numerical approximation. All these findings rule out the solution from \cite{Athens_2DBoussi,ChriChou_asymp,AsympNum_2D,ChertChriKurgan,ChriKolkVasi} as a candidate for the model of the real particles. The instability in time may be tackled by adding specially selected nonlinear terms that can either alleviate the blow-up or prevent the dissipation. The non-robustness with respect to the parameter giving the phase speed of the structure tells us that in order to make the standing localized wave moving, the profile behavior up to infinity has to be instantaneously changed.

\subsection{The Notion of a Pseudolocalized Solution}
The disappointment with the, so to say, ``strictly" localized solution requires that we reconsider the notion of localization. Since the equations that describe the field for a given family of particles are still under construction, one cannot be \emph{a-priori} sure what is the meaning of localization and what is the meaning of the primary variable. The requirement for strict localization that the profile function should be square integrable, while a closer look reveals that the quasi-particle behavior may be well satisfied if merely the Laplacian (the curvature) or the gradient of the profile (the slope) are square integrable. Such kind of solutions were first considered in  \cite{NonProbab} and called ``pseudolocalized". 

The idea of the pseudolocalized solution is that one solves a nonlinear equation for the strictly localized curvature and then integrate the Poisson equation with r.h.s being the known curvature to obtain the profile of the original amplitude. In 2D, such an integration will create a logarithmic behavior at infinity. Fortunately, in 3D the behavior of the solution of the Poisson equation is as $1/r$, which is very agreeable physically. Despite of the visual resemblance to a strictly localized solution, the profile is actually pseudolocalized in the sense of the above definition.

The ultimate question to answer is whether the 3D solitons can retain their shape when they are allowed to evolve according to time dependent equation. To demonstrate the presence of the collision property  numerically in 3D seems like an insurmountable task due to the required computational resources. The feasible goal is to first show that the shape of the soliton does not change when left to propagate according to the full equation. A prerequisite to investigating the collision properties of the pseudolocalized solutions is to find analytical or numerical representation of a single QP. This is the goal of the present paper.

\section{Selecting the model}

As already mentioned, we focus on the Boussinesq equation which arises in shallow water flows and in the theory of shells and plates. It is a generalized wave equation containing dispersion in the form of a biharmonic operator of the sought function. The dispersive effects of the biharmonic operator can be countered by the presence of a nonlinearity, and as a result a permanent wave of localized type my exist and propagate without change

Consider a generic version of Boussinesq equation
\begin{equation}
u_{tt} = \Delta [ u - \alpha_2 u^2 -\alpha_3 u^3 - \beta\Delta u]. \label{eq:BE}
\end{equation}
Alternatively, in shell theory the nonlinearity enters the picture in a different fashion (see the simplified version from \cite{NonProbab})
\begin{equation}
u_{tt} = \Delta u - \alpha_3 (\Delta u)^3 - \beta\Delta^2 u. \label{eq:nonprobab}
\end{equation}

To illustrate the idea, we introduce spherical coordinates
\begin{equation}
x = \rho \cos\theta \cos\phi,\quad y = \rho \cos\theta \sin\phi, \quad z = \rho \sin\theta
\end{equation} 
and  
restrict ourselves to the case of spherical symmetry when 
\begin{equation*}
\Delta : = \frac{1}{\rho^2} \DD{}{\rho} \rho^2 \DD{}{\rho}.
\end{equation*}

\section{The Linearized Equation and the notion of Pseudolocalized Solution}

The linearized version of the equation is merely the Euler--Bernoulli equation, i.e.,
\begin{equation}
u_{tt} = \Delta  u - \beta \Delta^2 u. \label{eq:Euler_Brnoulli}
\end{equation}

In 2D, the solution with point symmetry that does not exhibit a singularity at the origin was first found in different context (see, e.g.,  \cite{Lumini_WCNA}) to be
\begin{equation}
u^{(2)}=K_0(r/\sqrt{\beta}) + \ln (r/\sqrt{\beta}),
\end{equation}
where $K_0$ is the modified Bessel function of the second kind. Clearly, this solution is not localized function, because it increases to infinity at $r \rightarrow \infty$. What is is more important is that it is not even pseudolocalized, in the sense of the definition of the present work, because the following integral diverges (denote $\eta = \frac{r}{\sqrt{\beta}}$):
\begin{equation}
\int_\epsilon^R (u^{(2)}_r)^2 r \rd r = \int_\epsilon^R \big[ (K_0'(\eta))^2\eta + {2}K_0'(\eta)   + \frac{1}{\eta}\big] \rd \eta \propto const  - \ln \epsilon +  \ln R, \quad \text{for} \quad R\gg 1.
\end{equation}
Note that the integral diverges also at $\epsilon \rightarrow 0$.

In 3D, the solution of the linear problem with point symmetry that does not have a singularity at the origin is (see \cite{NonProbab})
\begin{equation}
u^{(3)}(\rho) = \frac{1 - \exp(-\rho/\sqrt{\beta})}{\rho}, \label{eq:spherical_symmetry}
\end{equation}
which decays at infinity, but is not localized, because the integral of the square of it diverges linearly at infinity.  However, it is a pseudolocalized solution because the $L^2$-integral of the gradient converges, namely
\begin{equation}
 \int_{0}^{\infty}  [\DD{u^{(3)}(\rho)}{\rho}]^2 \rho^2 \rd \rho = \int_{0}^{\infty}  \Big[-\frac{1 - \exp(-\rho/\sqrt{\beta})}{\rho^2} + \frac{\exp(-\rho/\sqrt{\beta})}{\rho\sqrt{\beta}}\Big]^2 \rho^2 \rd \rho  =\frac{1}{2\sqrt{\beta}}. 
\end{equation}

It is important to mention here that the above solution does not exhibit a singularity at the origin, but it is not a smooth function of the coordinates $(x,y,z)$.  It appears to be akin to the ``peakons" from \cite{CamassaHolm} (see, also the in-dept study of the numerical approach to peakons evolution \cite{ChertDuToitMarsden} )with the difference being that it decays at infinity as $1/\rho$ rather than $e^{-\rho}$. In addition, we have not been able to show that an analytical expression for the peakon can be obtained when the nonlinearity is not neglected. Actually, this paper deals with finding a numerical solution for the nonlinear case. An important advantage of the pseudolocalized solution is that if it propagates, then the additions to the profile is of order $1/r^2$, which does not qualitatively change the decay at infinity, as it is the case of strictly localized solutions. The profile stays in the the same class of localization. 

\section{Effect of Nonlinearity on the Shape of the Pseudolocalized Solution}

The most conspicuous trait of the above defined pseudolocalized solution is that it is not smooth at the origin of the coordinate system.  It is interesting to note that for some specific nonlinearity, such as the one in Eq.~\eqref{eq:nonprobab}, the effect of nonlinearity leads to smoothing of the profile at the origin (see \cite{NonProbab}), and the pseudolocailzed solution does not have the peak of the linear approximation. However, the way nonlinearity appears in Eq.~\eqref{eq:BE} does not act to smooth the peak. There is no known analytical solution for the nonlinear version of the governing equation, so we present here a numerical solution.

The full three-dimensional model requires significant computational resources. For this reason we have already restricted ourselves to the case of spherical symmetry. Now, the sought function depends only on the radial coordinate $\rho$. To emphasize this fact, we use a new variable $f(\rho) := u(x,y,z)$. Then the original equation is a equivalent to the following system
\begin{gather}
\beta  \frac{1}{\rho^2} \frac{\rd }{\rd \rho} \rho^2 \frac{\rd }{\rd \rho} f +\alpha_2 f^2 + \alpha_3 f^3 - f = g, \\
 \frac{1}{\rho^2} \frac{\rd}{\rd \rho} \rho^2 \frac{\rd}{\rd \rho} g = 0
\end{gather}

One can integrate the second of the above equations to obtain $g = C_1/\rho + C_2$. The trully localized solution requires that $C_1=C_2=0$. The pseudolocalized one is obtained when only $C_2 = 0$, and $C_1$ can have an arbitrary nontrivial value.  We use this fact to reduce the problem to solving numerically the single equation
\begin{equation}
\beta  \frac{1}{\rho^2} \frac{\rd }{\rd \rho} \rho^2 \frac{\rd }{\rd \rho} f+ \alpha_2 f^2 + \alpha_3 f^3 - f = \frac{a}{\rho}, \label{eq:1eq}
\end{equation}
where $a$ is at our disposal and choosing it will lead to different solutions. The boundary conditions are the behavioral condition at the origin, and the asymptotic condition at $\rho=\rho_\infty$ which requires that
\begin{equation}
\frac{\rd f}{\rd \rho} = \frac{1}{\rho_\infty} f,\label{eq:asymp_bc}
\end{equation} 
where $\rho_\infty$ is the finite value of the  radial coordinate at which the computational domain is truncated.

\subsection{Difference Scheme}

The boundary value problem  Eqs.~\eqref{eq:1eq}, \eqref{eq:asymp_bc}  is to be solved numerically.  We use a grid which is staggered by $\frac{1}{2}h$ from the origin $\rho=0$, while it coincides with the ``numerical infinity'', $\rho=\rho_\infty$. Thus
\[
\rho_i = \big(i-\tfrac{1}{2}\big)h,\quad \rho_{i\pm\frac{1}{2}}=\rho_i\pm\tfrac{1}{2}h, \quad i = 1, 2, \dots N,\]
\noindent where $h = {\rho_\infty}/(N- 0.5)$ and $N$ is the total number of points. 

The staggered grid gives a unique opportunity to create difference approximations for the Bessel operators involved in our model that take care of the singularities of the respective Bessel operator automatically, without the need to impose explicit behavioral boundary conditions in the origin. This is made possible by the fact that \(\rho_{\frac{1}{2}}= 0.\)

The main difference of the above and the smooth (non-peakon) solution is that the latter is a result of a bifurcation and exists only for one specific value of the amplitude of the soliton. The pseudolocalized solution exists for continuous interval of the parameter $a$ in Eq.~\eqref{eq:1eq}

The finite-difference complete  scheme  with the boundary conditions is given by the following system with tree-diagonal matrix:
\begin{subequations}
\begin{gather}
{\rho^2_{\frac{3}{2}}} \tilde f_{2} - \big[{\rho^2_{ \frac{3}{2}}} + h^2 \rho_1^2\big]\tilde f_1  
= - h^2 \rho^2_1 \big[ \alpha_2 \big(f^{n}_1\big)^2 + \alpha_3 \big(f^{n}_1\big)^3 \big] + a h^2 {\rho_1},\\
{\rho^2_{i + \frac{1}{2}}} \tilde f_{i+1} -\big[{\rho^2_{i - \frac{1}{2}}}+ {\rho^2_{i + \frac{1}{2}}}+h^2 \rho_i^2\big]\tilde f_i  + {\rho^2_{i - \frac{1}{2}}}\tilde f_{i-1}
= - h^2 \rho^2_i \big[ \alpha_2 \big(f^{n}_i\big)^2 + \alpha_3 \big(f^{n}_i\big)^3 \big] + a h^2 {\rho_i},
\nonumber \\
\quad i=2,\dots,N-1,\\
\tilde f_{N+1}-  \big(\frac{h}{\rho_N} +1)\tilde f_N =0.
 \label{eq:F_difference}
\end{gather}
\end{subequations}
Note that the first line is the so-called behavioral conditions stemming from the properties of the operator rather than imposting additional boundary condition.

To control the convergence we use also  relaxation
$ f^{n+1} := \omega \tilde f + (1-\omega) f^n,$ and repeat the iterations until convergence is attained when the value of the last iteration is taken as the solution for function $f$.

\subsection{Results and Discussion}

All numerical experiments to follow are carried out with $\beta=1$, since one can use a simple rescaling of the independent spatial variable. We set $\rho_\infty=1000$ and use $N=20001$ for the total number of grid points. In most of the cases reported below, a relaxation parameter $\omega=0.2$ was sufficient to secure convergence of the iterations. 

\begin{figure}[ht]
\centerline{\includegraphics{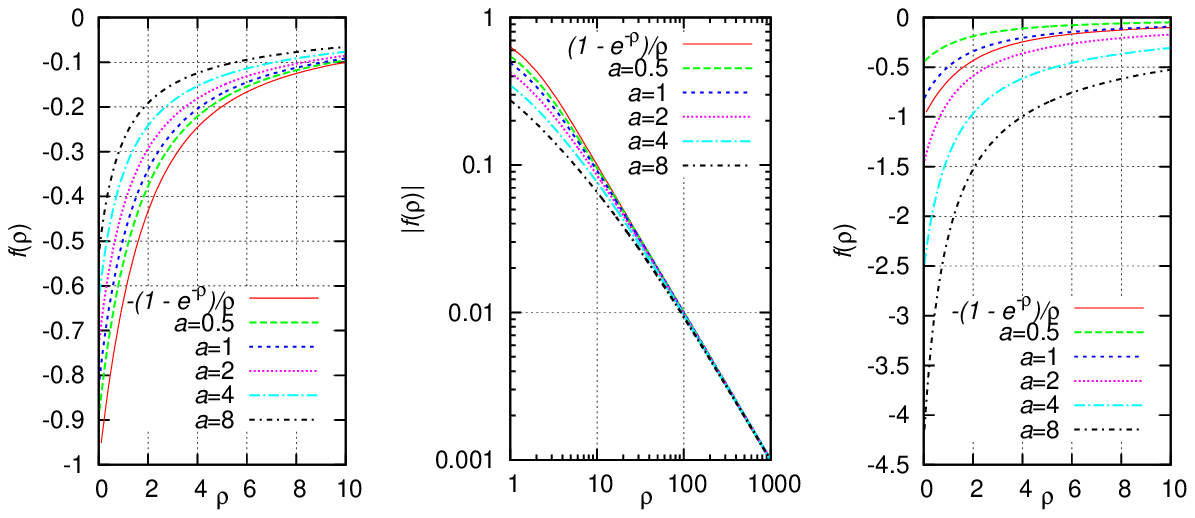}}
\caption{The peakon solution for the quadratic nonlinearity $\alpha_2=1$ and $\alpha_3=0$ and for different values of parameter $a$ from Eq.~\eqref{eq:1eq}. In the left and middle panel (log-log plot), the profile is scaled by $a$. In the right panel is the original profile.}\label{fig:quadratic}
\centerline{\includegraphics{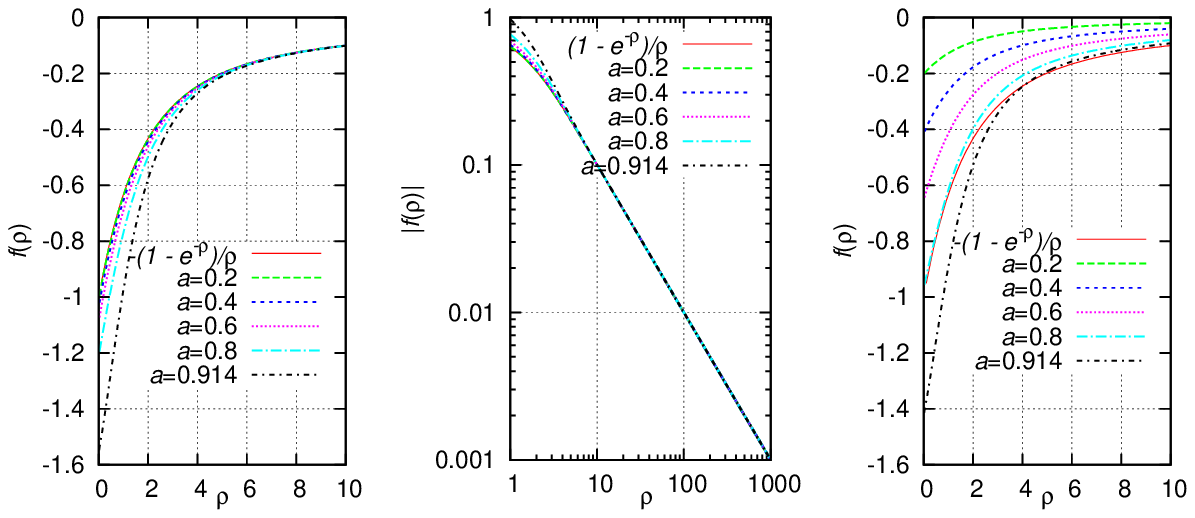}}
\caption{The peakon solution for the cubic nonlinearity $\alpha_2=0$ and $\alpha_3=1$ and for different values of parameter $a$ from Eq.~\eqref{eq:1eq}. In the left and middle panel, the profile is scaled by $a$. In the right panel is the original profile.}\label{fig:cubic_positive}
\end{figure}

First we examine the case of only quadratic nonlinearity, i.e., $\alpha_2=1$ and $\alpha_3=0$. The interesting observation is that the algorithm is not convergent for negative $a$ when the profile becomes positive. For positive $a$, the profile becomes taller with the increase of $a$ (see the rightmost panel of Fig.~\ref{fig:quadratic}), but the increase is lesser than the increase of the coefficient $a$ as seen from the left  panel of Fig.~\ref{fig:quadratic}, where the profile is scaled by $a$. The middle panel shows that for very large distance from the origin, all of the scaled profiles collapse  on the linear solution.
One can say that the nonlinearity acts to reduce the overall amplitude of the solution, all other conditions equal.

In a similar fashion we compute the numerical solution for the cubic case $\alpha_2=0$ and $\alpha_2\ne 0$. It is well known that for the 1D cubic Boussinesq equation, the case $\alpha_3 >0$ does not lead to localized solutions. More specifically, for $\alpha_3 >0$ solitons of $sech$-type cannot exist.  It is interesting to note that we actually found pseudolocalized solutions even for this case, but only for $a \le 0.914$. In fact, for the last case the solution was obtained only when we reduced the relaxation parameter to $\omega =0.005$ and increased the number of grid points to $40001$, all other parameters being the same.  However, all our efforts to obtain a result for $a= 0.915$ did not meet with a success. We also mention here that for cubic nonlinearity, the positive and negative amplitudes are equally possible for the solution, and we present here the case of $a >0$ which results in negative amplitude of the peakon. The results are presented in Fig.~\ref{fig:cubic_positive}.
Now, as it should have been expected, the profile is lower than the linear one for all values of $a$ (see the right panel of  Fig.~\ref{fig:cubic_positive}). Yet, the nonlinearity retains its steepening effect (see the left panel of  Fig.~\ref{fig:cubic_positive}.). Naturally, because of the smaller values of $a$, the steepening effect is less obvious.

\begin{figure}[h]
\centerline{\includegraphics{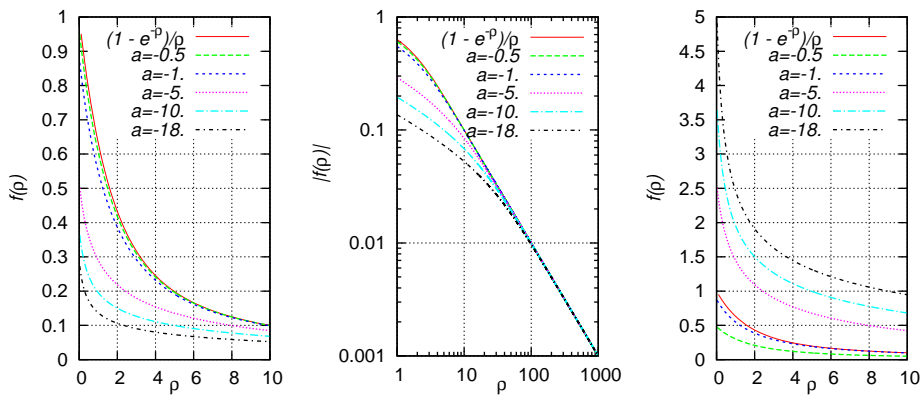}}
\caption{The peakon solution for the cubic nonlinearity $\alpha_2=0$ and $\alpha_3=-1$ and for different values of parameter $a$ from Eq.~\eqref{eq:1eq}. In the left and middle panel, the profile is scaled by $a$. In the right panel is the original profile.}\label{fig:cubic}
\end{figure}

Finally, we treat the cubic case with $\alpha_3<0$, when the energy functional is not positive definite allowing the bifurcation and appearance of 1D $sech$ solutions in the 1D case of strictly localized solutions. We found solutions with very large $a$. With $\omega=0.2$ we easily reached $a=18$, but for larger $a$ the relaxation parameter must be smaller. The result suggests that eventually,  the solution may exist for arbitrary $a$, but to claim this, one needs an analytical result.  Once again, both positive and negative solutions are possible and are equivalent upon changing the sign of the amplitude (or what is the same: the sign of $a$). Fig.~\ref{fig:cubic} presents the result for negative $a$, when the solution has a positive amplitude. 
Qualitatively, it is very similar to the quadratic case, with the small quantitative difference that the profiles are now steeper for the same $a$ than in the quadratic case.

\section{Conclusions}

In the present work we have argued the need to look beyond the strictly localized solutions when the concept of Quasi-Particle (QP) is applied. We consider the Boussinesq equation as the featuring model, and introduce the concept of \emph{pseudolocalized} solution which does not belong to the $L^2$-space itself, but its gradient and Laplacian do. It turns out that the pseudolocalized solution exists only in dimension three or greater. A similar attempt in 2D produces a clearly no-localized profile which  behaves at infinity as $\ln r$, where $r$ is the polar coordinate. Its gradient behaves as $1/r$, which means that it is not localized either. Thus the 2D solution is not  pseudolocalized.

For the 3D case we found a pseudolocalized solution for the linearized version of the equation and showed that its profile decays at infinity as $1/\rho$ (where $\rho$ is the spherical radial coordinate), which is rather different from the exponential decay of the strictly localized solution.
We shown that the profile is not smooth at the origin of the coordinate system, and can be called ``radial peakon''.

We devised a difference scheme to solve the full nonlinear equation numerically and showed that retaining the nonlinearity does not change qualitatively the profile obtained from the linear equation. Naturally, at infinity the profile decays with the same law $1/\rho$, since the nonlinear solution gradually approaches the linear one. The nonlinearity plays a role in steepening the peak of the solution at the origin.

The importance of this new kind of solution brings the concept of quasi-particle a step closer to modeling real particles in physics as localized solutions of a field equation (see the pioneering works of Skyrme \cite{Skyrme61,Perring_skyrme}), since the potential of the interaction of two QPs is defined by the decay of their ``tails". Since the decay here is $1/\rho$, the predominant part of the asymptotic behavior of the potential of interaction will be close to $1/\rho$ which yields to a gravitation-like force of interaction between the radial peakons. The results of the present work may prove instrumental in a new line of investigation in which the QPs are represented by the radial peakons investigated here.

{\bf Acknowledgment:}  This work is supported in part by the National Science Fund of Ministry of Education, Science, and Youth of Republic Bulgaria under grant DDVU02/71. The author is indebted to the anonymous referees for helpful suggestions.

\bigskip
{\bf References}

%\bibliography{../pseudolocalized}

\end{document}